\documentstyle[psfig]{mn}

\newcommand{\HI}{H\,{\sc i}}

\newcommand{\Ha}{H$\alpha$}
\newcommand{\kms}{~km\,s$^{-1}$}
\newcommand{\kkms}{km\,s$^{-1}$}

\newcommand{\FHI}{$F_{\rm HI}$}

\newcommand{\MHI}{$M_{\rm HI}$}

\newcommand{\Msun}{$M_{\odot}$}

\newcommand{\Lsun}{~L$_{\odot}$}
\newcommand{\Ho}{H$_{\rm o}$}

\title[A Wide-Field \HI\ Study of the NGC 1566 Group]
      {A Wide-Field \HI\ Study of the NGC 1566 Group\thanks{The 
       observations were obtained with the Australia Telescope which 
       is funded by the Commonwealth of Australia for operations as 
       a National Facility managed by CSIRO.}}

\author[V. A.~Kilborn et al.]
       {Virginia A. Kilborn$^{1,2}$,
        B\"arbel S. Koribalski$^2$, 
        Duncan A. Forbes$^1$,
\newauthor
        David G. Barnes$^3$,	
        Ruth C. Musgrave$^1$ \\
	$^1$Centre for Astrophysics \& Supercomputing, Swinburne University of Technology,  Mail 31, PO Box 218, Hawthorn, VIC 3122, Australia\\
	$^2$Australia Telescope National Facility, CSIRO, 
	    P.O. Box 76, Epping, NSW 1710, Australia\\
	$^3$School of Physics, University of Melbourne,  Parkville, VIC 3010, Australia}

\date{Received date; accepted date}

\begin{document}

\maketitle

\begin{abstract}
We report on neutral hydrogen observations of a $\sim5.5\degr\times
5.5\degr$ field around the NGC~1566 galaxy group with the multibeam
narrow-band system on the 64-m Parkes telescope. We detected thirteen
\HI\ sources in the field, including two galaxies not previously known
to be members of the group, bringing the total number of confirmed
galaxies in this group to 26. Each of the \HI\ galaxies can be
associated with an optically catalogued galaxy. No 'intergalactic \HI\
clouds' were found to an \HI\ mass limit of $\sim 3.5 \times 10^8 $
\Msun. We have estimated the expected \HI\ content of the late-type
galaxies in this group and find the total detected \HI\ is consistent
with our expectations. However, while no global \HI\ deficiency is
inferred for this group, two galaxies exhibit individual \HI\
deficiencies. Further observations are needed to determine the gas
removal mechanisms in these galaxies.

\end{abstract}

\begin{keywords}
   surveys,
   radio lines: galaxies,
   galaxies: individual (NGC~1566, NGC~1533, NGC~1549, NGC~1553),  
   galaxies: clusters: general.
\end{keywords}
 
\section{Introduction} 

\begin{figure*} 
\begin{tabular}{c}
\mbox{\psfig{file=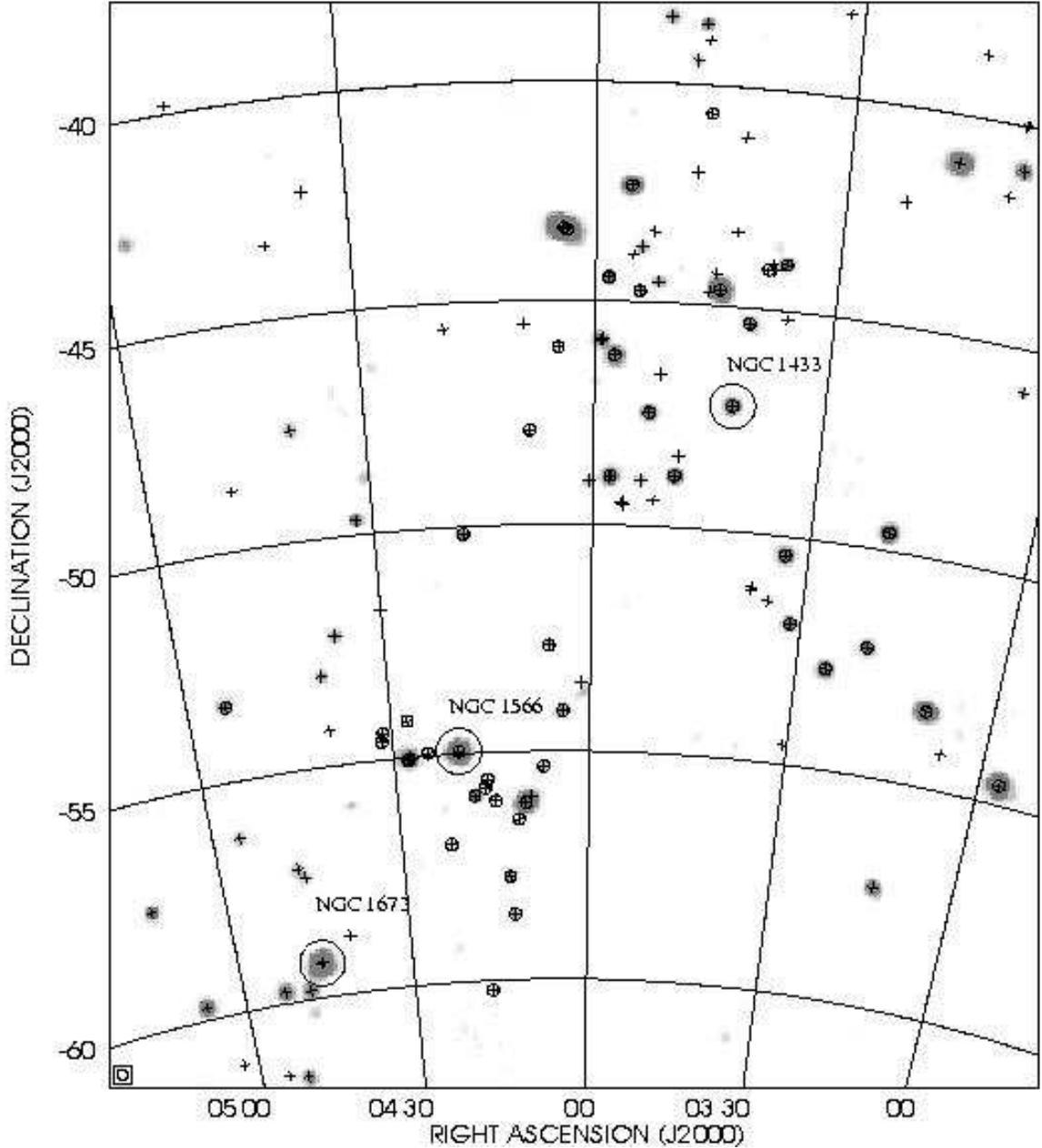,height=17cm}}
\end{tabular}
\caption{The \HI-rich galaxies in the Dorado group as revealed by 
   HIPASS. The measured \HI\ flux
   densities are shown by the grey scale. The gridded beam is
   15\farcm5, and is displayed in the lower left corner. The crosses
   denote all galaxies in the velocity range from 500 to 2400\kms\
   recorded in NED, and the open circles denote the 46 members of the
   Dorado Group as listed by Maia et al. (1989). Large circles mark
   the galaxies NGC~1433, NGC~1566, and NGC~1672, which dominate the
   smaller groups within the Dorado group.}
\label{fig:hipmom0}
\end{figure*}


The majority of galaxies in the Universe reside in galaxy groups (e.g.
Tully 1987). However the physical processes operating in groups are
poorly understood, and no detailed census of their contents is
available. Studies of neutral hydrogen in groups have previously
concentrated almost exclusively on Hickson Compact Groups (e.g.
Shostak, Allen \& Sullivan 1984; Verdes-Montenegro et al. 2001), as
they require small observing regions. Studies of loose groups, which
are more extended, are much fewer in number. The advent of the
multibeam receiver on the Parkes telescope (Staveley-Smith et
al. 1996) has made the study of loose groups feasible, and the success
of the instrument has been shown with new group members being found in
\HI\ surveys of optically well studied groups. For example, ten new
members were found in the Centaurus\,A group as part of the \HI\
Parkes All Sky Survey, HIPASS (Banks et al. 1999) and new dwarf
galaxies were found in \HI\ surveys of the NGC 5044 and NGC 1052 groups
(McKay et al. 2004). Several new galaxy groups were identified in the
HIPASS Bright Galaxy Catalogue (BGC, Koribalski et al. 2004).

\HI\ surveys of groups can reveal the effects of tidal interactions in
these environments (Haynes, Giovanelli \& Chincarini 1984). One of the
best examples of this is the VLA survey of the M81 group by Yun et
al. (1994), where the \HI\ distribution encompasses the major galaxies
in the group with many streams and tidal tails. Further tidal tails
were observed when the M81 group was observed during the \HI\ Jodrell
All Sky Survey, HIJASS (Boyce et al. 2001). Blind \HI\ surveys of
groups can turn up unexpected results. For example, Barnes \& Webster
(2001) imaged the \HI\ environment in five nearby galaxy groups,
leading to the discovery of a ring of \HI\ gas encircling the compact core
of the loose group LGG 138. \HI\ surveys can also uncover information
about galaxy formation and evolution in groups. This was the case when
a large bridge of \HI\ was discovered with Australia Telescope Compact
Array (ATCA) mapping of the NGC 6221/6215 galaxy group (Koribalski \&
Dickey 2004).

Cold Dark Matter theory predicts many more low-mass galaxies
surrounding large galaxies than are actually observed in groups
(e.g. Moore et al. 1999, Klypin et al. 1999). Searches for these
low-mass satellites have been made in \HI\ in several nearby groups
such as Centaurus\,A and Sculptor (de Blok et al. 2002), NGC 5798,
5962, 5970, 6278, 6500 and 6574 to a limit of $\sim7\times 10^6
$\Msun\ (Zwaan 2001), and NGC 1808 (Dahlem et al. 2001). More
recently, Pisano et al. (2004) surveyed three groups (LGG 93, LGG 180
\& LGG 478), similar in size and structure to the Local Group, for
low-mass \HI\ clouds (to an \HI\ mass sensitivity of $\sim 10^7
$\Msun). No population of \HI\ rich satellites with little, or no
optical emission were uncovered in any of the above surveys, leading
to the conclusion that massive \HI\ clouds are rare in the group
environment. However, to date, no \HI\ surveys have been deep enough
to rule out a population of low-mass ($< 10^7$ \Msun) \HI\ clouds.

Pointed \HI\ surveys of groups and clusters of galaxies have generally
found a trend for galaxies to be \HI\ deficient as a function of
distance from their centre. Verdes-Montenegro et al. (2001) found in a
homogeneous survey of 48 Hickson Compact Groups (HCGs) that they
contained only 40\% of the expected \HI\ mass, based on the optical
properties of the individual galaxies. Solanes et al. (2001) analysed
the \HI\ properties of 18 clusters, and found that two-thirds of their
sample were \HI\ deficient in the central regions (within the Abell
radius). However, a recent study of the \HI\ content of compact groups
of galaxies (Stevens et al. 2004) has found no strong evidence for an
\HI\ deficiency in their sample (15 groups). Their study highlights
the large uncertainties inherent in determining the expected \HI\ mass
of a galaxy based on its optical properties.

We are conducting wide-field \HI\ observations as part of the Group
Evolution Multiwavelength Study (GEMS) (Osmond \& Ponman 2004,
hereafter OP04; McKay et al. 2004), containing $\sim 60$ groups
selected to have existing ROSAT PSPC X-ray observations. OP04 detail
the selection criteria and X-ray properties of the groups.  Seventeen
of these groups in the southern hemisphere were surveyed for \HI\ with
the multibeam narrow-band system on the 64-m Parkes telescope. The
main aims of this \HI\ survey are to provide a census of the \HI\ gas
in groups, to find new group members and possible intergalactic \HI\
clouds, and to make a direct comparison between hot and cold gas in
the group environment for the first time. Here we report on our \HI\
results for the NGC 1566 group.

\subsection{The NGC~1566 Galaxy Group}

NGC 1566 is the brightest spiral member of a nearby group of galaxies
in Dorado at a distance of about 21 Mpc (OP04). This
distance, and an \Ho\ = 70\kms\,Mpc$^{-1}$ is used throughout the
paper. The Dorado group (see Fig.~1) consists of at least 46 galaxies
(Maia et al. 1989) and covers a velocity range from $\sim$500 to
2000\kms. It is part of the Fornax Wall which connects the NGC~1672,
NGC~1566 and NGC~1433 galaxy groups.

Figure~\ref{fig:hipmom0} shows the \HI\ distribution of the Dorado
group. The positions of known optical galaxies are marked.  This map
was derived from the
HIPASS\footnote{http://www.atnf.csiro.au/research/multibeam} data
cubes which have a velocity resolution of 18\kms\ (see Barnes et
al. 2001). The {\sc AIPS} task {\sc momnt} was used with Hanning
smoothing over five channels (65\kms), five pixels (20\arcmin) and a
flux density cutoff, after smoothing, of 20 mJy\,beam$^{-1}$ per
channel. The \HI\ distribution of the region shows three distinct
groupings of galaxies around the galaxies NGC 1433, NGC 1672 and NGC
1566. These three groups are at similar velocities and thus are part
of the larger complex.

The NGC 1566 group has been the subject of several optical studies.
Huchra \& Geller (1982) catalogue 18 members in the group (which they
call HG3). The group catalog by Garcia (1993) lists 6 members, and
they derive a group recession velocity of 1292 \kms\ and a velocity
dispersion of 99 \kms.  Ferguson \& Sandage (1990), Morshidi-Esslinger
et al. (1999) and Carrasco et al. (2001) surveyed various areas of the
Dorado group in the optical and cataloged large numbers of mostly low
luminosity dwarf galaxies as potential members. The LEDA database
lists 24 galaxies with known velocities that lie within the NGC 1566
group, while the NASA Extragalactic Database (NED) lists 22 previously
catalogued galaxies in the region.

Several members of the NGC 1566 group have been previously mapped in
\HI\ at the ATCA. Walsh (2004) studied the \HI\ dynamics of the
spiral galaxy NGC 1566 itself, finding a nearly circular \HI\
envelope, and a total dynamical mass of $1.2 \times
10^{11} $\Msun. Many other observations have been made of this object,
including CO (Bajaja et al. 1995), \Ha\ (Pence, Taylor \& Atherton,
1990), \HI\ (Reif et al. 1982), and X-ray and radio continuum (Ehle et
al. 1996). NGC 1533 is a nearby, early-type galaxy with two small
companions, IC\,2038/9. Recent ATCA observations obtained by
Ryan-Weber, Webster \& Bekki (2002) reveal an asymmetric \HI\ ring
surrounding the optical galaxy, with only a small amount of \HI\
associated with IC\,2038 and none with IC\,2039. The interacting pair
of galaxies NGC 1596/1602 has recently been mapped in \HI\ at the
ATCA (Chung et al. 2004).

OP04 studied the X-ray properties of the NGC~1566 group using ROSAT
PSPC observations. They found galaxy halo emission around NGC~1566
itself, (log $L_{\rm X}$ = 40.41 erg s$^{-1}$, see their Table~4), but
no group-scale emission. This suggests that the group is relatively
young and not yet virialised. The NGC~1566 group centre and radius, as
defined in the GEMS sample (see OP04, their Table~1), is
$\alpha,\delta$(J2000) = $04^{\rm h}\,20^{\rm m}$\,00\fs6,
--54\degr\,56\arcmin\,17\arcsec\ and $r_{500}$ = 0.47 Mpc, where
$r_{500}$ is the radius at 500 times the critical density of the
Universe at the current epoch. OP04 derive a group velocity of $v =
1402 \pm 61$\kms\ and a velocity dispersion of $\sigma_{\rm v} = 184
\pm 47$\kms, based on the nine group members within $v \pm
3\sigma_{\rm v}$ and $r_{500}$. The total $B$-band luminosity of the
NGC~1566 group based on these nine members is $1.87 \times
10^{11}$\Lsun.

In this paper we present our \HI\ survey of the NGC 1566 group. In
Section~2 we summarize the observations and data reduction, along with
the source detection and optical identification. In
Section~3 we give the results from the \HI\ survey. Discussion of the
results is provided in Section~4, including the optical properties and
\HI\ content of the group, X-ray properties, and dynamics.  Finally we
present our conclusions in Section~5.

\section{Observations and data reduction} 

\begin{table} 
\centering
\caption{Narrow-band observing and cube parameters.}
\label{tab:table2} 
\begin{tabular}{lc}
\hline
Gridded beam size      & 15\farcm5\\
Total observing time    & 18.5 hr \\
Velocity range          & 400 -- 2080\kms  \\
Channel width           & 1.65 \kms \\
Velocity resolution     & 2.6 \kms \\
rms noise per channel   & 18.8 mJy \\
\hline
\end{tabular}
\flushleft
\end{table}

\begin{table*} 
\caption{\HI\ properties of the detected galaxies.}
\label{tab:hi_params} 
\begin{tabular}{lcrrrrrrrrrrr}
\hline
No.  & $\alpha,\delta$(J2000)&$v$& $w_{50}$& $w_{20}$& S$_{\rm peak}$ & \FHI &rms & Order & Box & \MHI \\
         & [$^{\rm h\,m\,s}$], [\degr\,\arcmin\,\arcsec]
          & [\kkms] & [\kkms] & [\kkms] & [Jy] & [Jy\kms] & [Jy]&  &  & [$10^8 $\Msun] \\
(1) & (2) & (3) & (4) & (5) & (6) & (7) & (8) & (9) & (10)& (11)\\
\hline

1 & 4:12:11,$-$58:34:17 & 1466$\pm$ 3&  55$\pm$   6&  102$\pm$  9& 0.106$\pm$0.009  &   5.7$\pm$  0.7&   0.008&  3&    5 &   5.9$\pm$  0.7 \\
2 & 4:10:51,$-$56:30:32 & 1310$\pm$ 7& 104$\pm$  14&  129$\pm$ 21& 0.028$\pm$0.007  &   2.0$\pm$  0.6&   0.006&  3&    5 &   2.1$\pm$  0.6 \\
3 & 4:12:39,$-$57:45:50 & 1176$\pm$ 3& 189$\pm$   6&  214$\pm$  9& 0.076$\pm$0.008  &   8.7$\pm$  0.9&   0.004&  3&    5 &   9.0$\pm$  0.9 \\
4 & 4:27:25,$-$57:27:21 & 1215$\pm$ 7&  75$\pm$  14&  137$\pm$ 21& 0.048$\pm$0.007  &   3.2$\pm$  0.6&   0.005&  3&    5 &   3.3$\pm$  0.6 \\
5 & 4:09:42,$-$56:06:55 &  796$\pm$ 3& 247$\pm$   6&  324$\pm$  9& 0.382$\pm$0.020  &  73.7$\pm$  2.9&   0.012&  3&    13&  76.5$\pm$  3.0 \\
6 & 4:22:42,$-$56:16:59 & 1350$\pm$ 4&  52$\pm$   8&   76$\pm$ 12& 0.049$\pm$0.007  &   2.4$\pm$  0.5&   0.005&  7&    5 &   2.5$\pm$  0.6 \\
7 & 4:14:40,$-$56:04:25 & 1298$\pm$ 5& 189$\pm$  10&  343$\pm$ 15& 0.140$\pm$0.010  &  23.2$\pm$  1.3&   0.011&  3&    13&  24.1$\pm$  1.4 \\
8 & 4:17:56,$-$55:57:06 & 1370$\pm$ 2& 193$\pm$   4&  218$\pm$  6& 0.124$\pm$0.009  &  16.4$\pm$  1.1&   0.004&  3&    5 &  17.1$\pm$  1.2 \\
9 & 4:07:10,$-$55:18:01 & 1068$\pm$ 3&  99$\pm$   6&  119$\pm$  9& 0.073$\pm$0.008  &   5.9$\pm$  0.7&   0.005&  3&    5 &   6.1$\pm$  0.8 \\
10& 4:19:53,$-$54:56:46 & 1502$\pm$ 1& 200$\pm$   2&  223$\pm$  3& 1.040$\pm$0.052  & 148.1$\pm$  6.4&   0.009&  3&    13& 153.9$\pm$  6.7 \\
11& 4:27:45,$-$55:01:08 & 1572$\pm$ 3&  93$\pm$   6&  174$\pm$  9& 0.283$\pm$0.016  &  28.1$\pm$  1.6&   0.009&  3&    11&  29.2$\pm$  1.7 \\
12& 4:03:56,$-$54:05:04 & 1180$\pm$ 2& 357$\pm$   4&  381$\pm$  6& 0.101$\pm$0.009  &  15.9$\pm$  1.1&   0.004&  3&    5 &  16.5$\pm$  1.2 \\
13& 4:05:42,$-$52:41:01 &  902$\pm$ 2&  88$\pm$   4&  103$\pm$  6& 0.079$\pm$0.008  &   4.8$\pm$  0.6&   0.007&  3&    5 &   5.0$\pm$  0.7 \\

\hline
\end{tabular}
\flushleft The columns are (1) GEMS galaxy number, (2) fitted \HI\
centre position, (3) heliocentric velocity in the optical convention,
(4) 50\% velocity width, (5) 20\% velocity width, (6) \HI\ peak flux
density, (7) integrated \HI\ flux density, (8) clipped rms per channel
(9) Order of baseline fit to the \HI\ spectrum, (10) Box size used to
create the \HI\ spectrum in 4\arcmin\ pixels, (11) \HI\ mass for the
detection assuming cluster distance of 21 Mpc. The errors are derived
following Koribalski et al. (2004).
\end{table*}

Our observations of the NGC~1566 group follow the approach described
in McKay et al. (2004).  We scanned a field of dimensions $5.5^\circ
\times 5.5^\circ$ using the Parkes 64~m radiotelescope with the 21~cm
Multibeam Receiver \cite{staveley1996} installed at the prime focus.
Observations were acquired in the periods 2002 January 11--18, 2002
March 8--9 and 2003 May 3--5.  The field was centred at $\alpha,
\delta$ (J2000) =  04$^{\rm h}$16$^{\rm m}$00$^{\rm s}$,
$-$55$\degr$35$\arcmin$00$\arcsec$.  Scans were made at a rate of one
degree per minute, along lines of equatorial latitude and longitude
separated by 4~arcmin. A total of 198 scans were acquired.

Data were reduced using {\sc livedata}, a component of the {\sc
aips++} package.  On input, raw correlator spectra were convolved with
a Hanning filter to eliminate ringing from narrow and bright emission
sources.  Subsequent processing steps were based on those for
HIPASS data \cite{barnes01} and previous GEMS data
\cite{mckay2004}, with changes made to improve dynamic range near
bright sources, and to track the time evolution of the (source-free)
sky at higher order.  Bandpass correction was applied on a per-beam,
per-scan basis by iteratively clipping the data and fitting a 2nd
degree polynomial to the time series of each channel.  Data acquired
within 20~arcmin of known \HI\ sources in the field (based on a first,
quick process and search step) were masked during fitting to prevent
contamination of the calculated bandpass.  Each bandpass-removed
spectrum was frame-shifted to the barycenter of the Earth-Sun system,
and then baselined by subtracting a source-masked, iterative, clipped
2nd-degree polynomial fit, this time in the spectral domain.

Calibrated, processed spectra were imaged using the {\sc gridzilla}
component of {\sc aips++}, which is described in Barnes et al. (2001).
We used the {\sc wgtmed} statistic, which calculates mesh pixel values
by taking the weighted median of data within 6~arc\-min of the centre
of each pixel.\footnote{For comparison, HIPASS images use the {\sc
gridzilla} {\sc median} statistic which is the median of weighted
values} The weight values were directly proportional to the canonical,
Gaussian observing beam profile which has a FWHM of $14.4$~arc\-min.
We produced 2026 channel maps, centred on $V_{\rm CMB} = 1231.8$~\kms,
with velocity widths of $0.824$~\kms. 

The gridded cube was smoothed in the spectral domain again with a
Hanning filter three channels wide, and every second channel was
discarded. The velocity resolution of the image was then measured,
yielding a FWHM spectral resolution of $2.6$~\kms. The channel
separation in the final cube is 1.65\kms.  The rms noise in the final
cube is measured to be 18.8 mJy per channel. The final gridded
beamsize is 15.5 arcminutes, and the pixel size is 4\arcmin\
square. The final cube parameters are summarised in
Table~\ref{tab:table2}.

\begin{figure*} 
\centering
\psfig{file=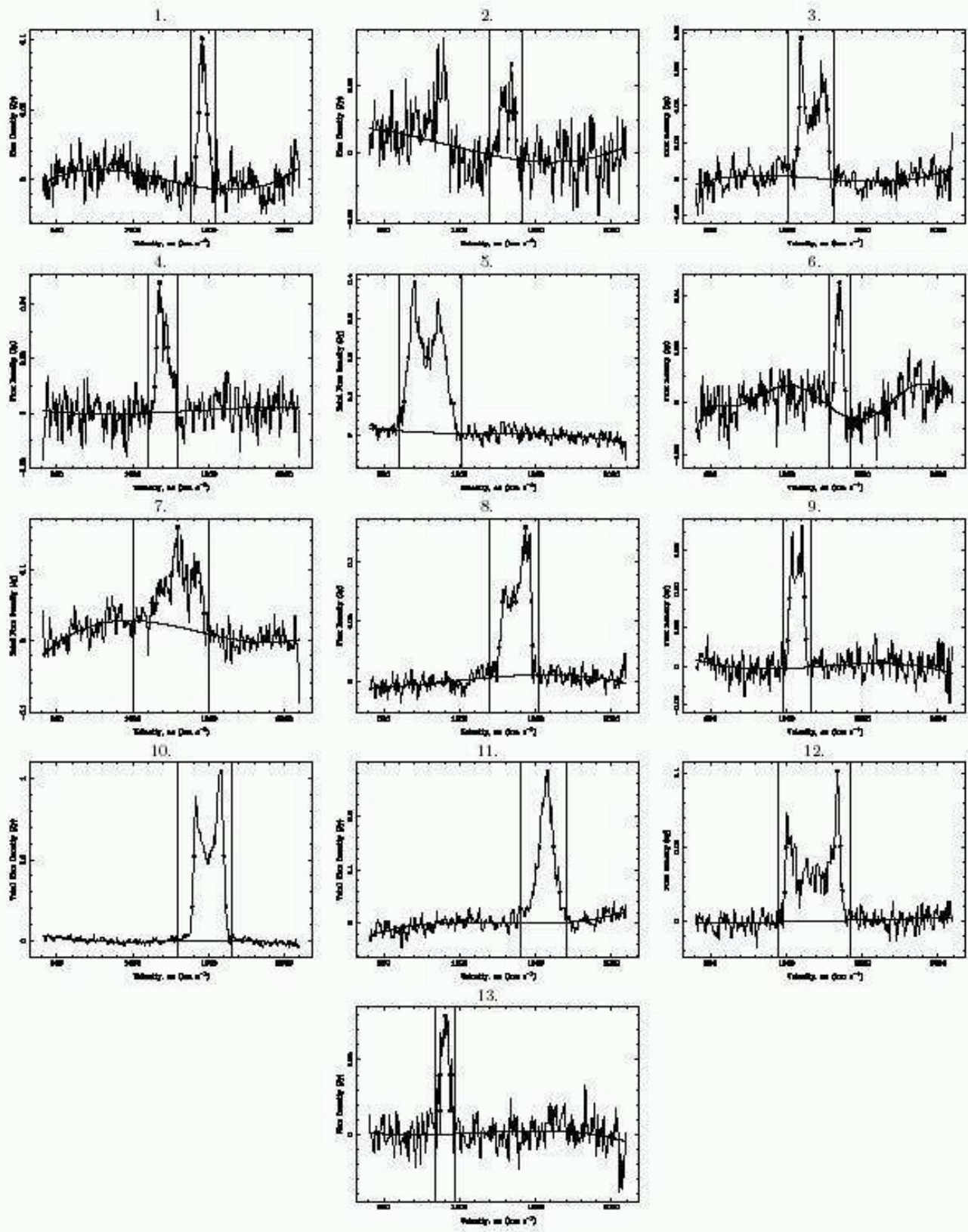,height=22cm}

\caption{\HI\ spectra of the thirteen galaxies detected in our
   NGC~1566 survey. The spectra were Hanning smoothed to a velocity
   resolution of 6.6\kms. The fitted baseline is shown and the \HI\
   peak flux density is marked with a filled circle. The $w_{20}$ and
   $w_{50}$ velocity widths are shown by the open circles (outer fit),
   and crosses (inner fit). The velocity region between the vertical
   lines in the spectra was disregarded in the baseline fit.}
\label{fig:hispectra}
\end{figure*}

\begin{figure*} 
\centering
\begin{tabular}{c}
\mbox{\psfig{file=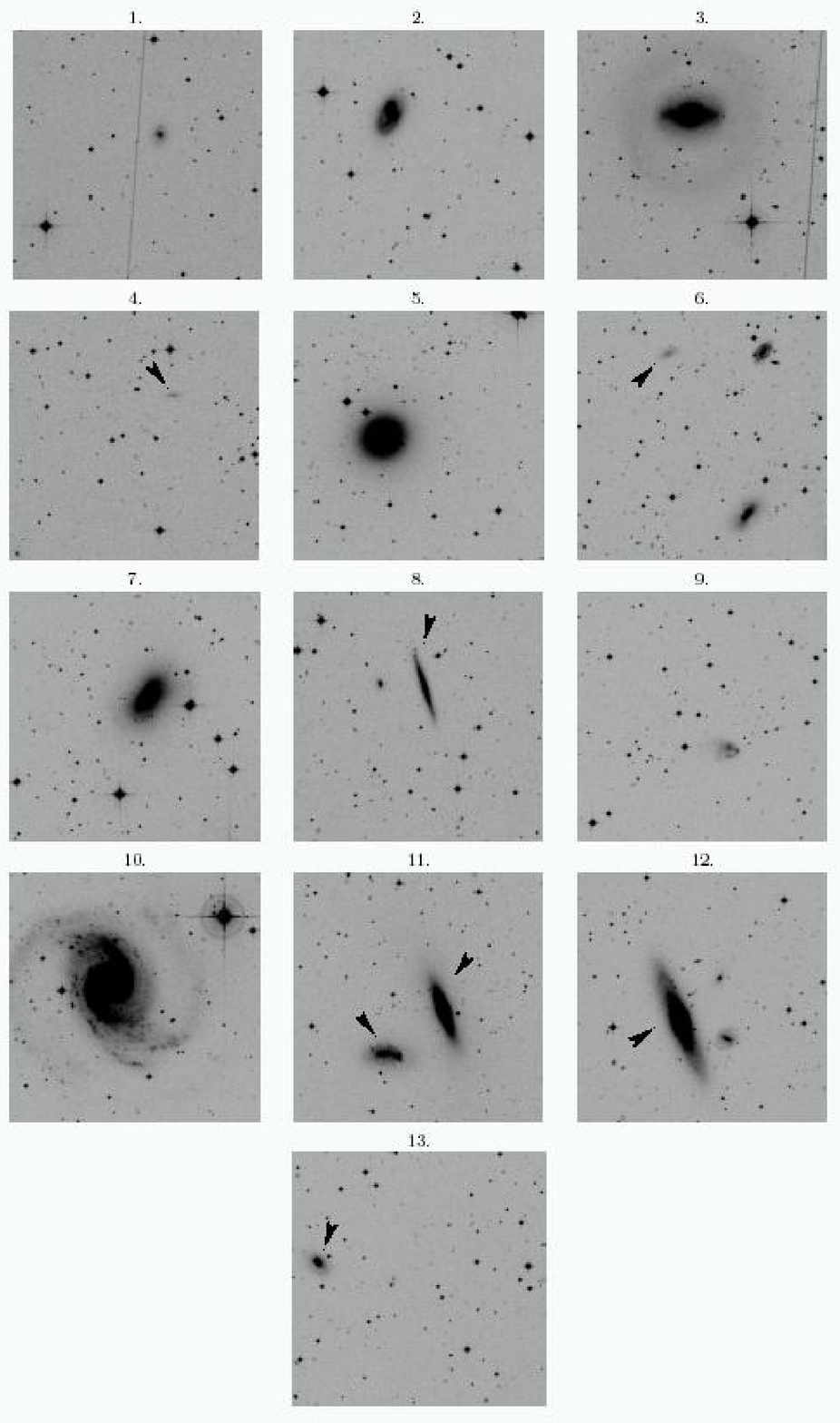,height=22cm}}\\
\end{tabular}

\caption{Second generation DSS $R$-band images of the thirteen galaxies
   detected in the NGC~1566 group. Each image is 10 arcmins across,
   and is centred at the \HI\ detection. When there is more than one
   galaxy in the field of view, arrows indicate the matching optical
   counterpart, determined by either the known optical velocity of the
   galaxy, or in the case of \#4 and \#6, from high resolution ATCA
   observations. Note that in fields \#8 and \#12  there are
   several galaxies within the field of unknown redshift.}
\label{fig:dssimages}
\end{figure*}

\subsection{Source Finding}

The \HI\ data cube was searched visually for sources using the {\sc
kview} visualisation program (Gooch 1995). Two of us (VAK \& RCM)
catalogued sources from the original cube, and also from two smoothed
versions of the cube, with final velocity resolution of 3.3\kms\ and
6.6\kms. The rms in the smoothed cubes was 11 mJy beam$^{-1}$ and
$\sim$ 8 mJy beam$^{-1}$ respectively. To determine the detection limit
for sources in the cube, and as an extra check for the searching,
twenty fake point sources were injected randomly into the original
cube. These sources had a range of peak flux density between 10$-$30
mJy, and velocity width of 50$-$500 km s$^{-1}$. Two fake-injected
cubes were made, to ensure if one of the randomly placed fake sources
lay upon a real \HI\ signal, the other cube would likely have it
unobscured. Both the unsmoothed, and two smoothed versions were
searched for sources, as with the original datacube. From the recovery
rate of fake sources, we have determined our survey to be complete to
a peak flux density limit of 58 mJy ($\approx 3\sigma$) in the
unsmoothed cube. While we did detect a number of sources below this
limit, the sources with the lowest peak flux density were detected in
the smoothed cubes only. No relationship with velocity width was seen
with the detectability of these fake sources. Assuming a galaxy
velocity width of 100 \kms\ and a Gaussian \HI\ profile, this translates to
an \HI\ mass limit of $3.5 \times 10^8 $\Msun\ ($\sim 1 \times
10^9 $ \Msun\ for a velocity width of 300 \kms).

\begin{figure*}
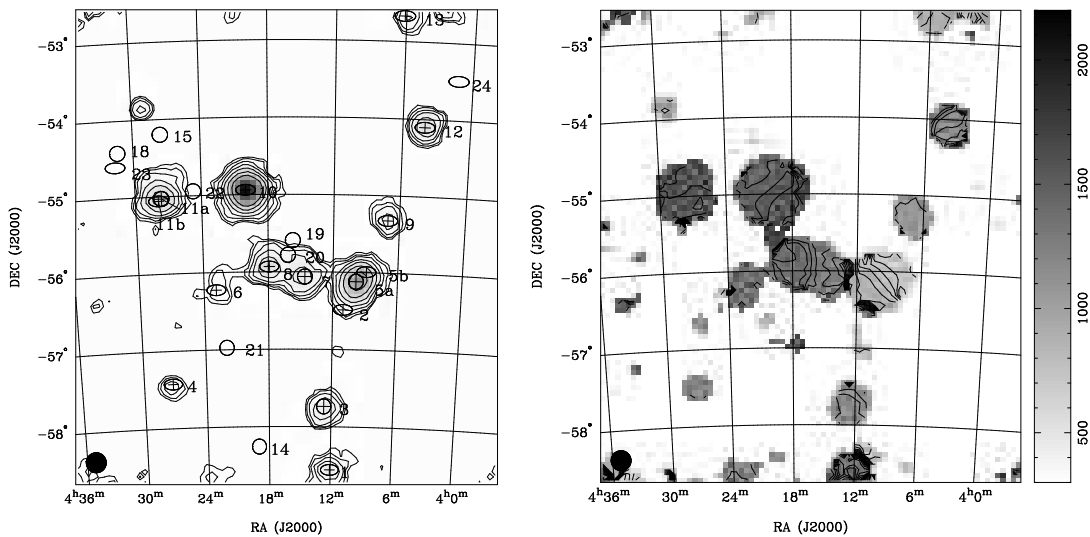
 %
\begin{tabular}{cc} 
\mbox{\psfig{file=kilbornva_5_a.ps,height=7cm,angle=-90}} &
\mbox{\psfig{file=kilbornva_5_b.ps,height=7cm,angle=-90}}
\end{tabular}
\caption{(a) The integrated \HI\ intensity map of the NGC~1566 group
   as obtained from the \HI\ Parkes narrow-band observations.  The
   measured \HI\ flux densities (contours) for the left image are 0.5,
   1, 2, 4, 8, 16 and 32 Jy beam$^{-1}$ \kms.  Overplotted are the
   positions of known group members. The 13 crossed symbols are those
   detected in our \HI\ survey, and the open symbols were undetected
   in \HI. The circles represent E/S0 galaxies, and ellipses represent
   late-type galaxies. The numbers correspond with those in
   Table~\ref{tab:opt_hi}. Note that galaxies 16 \& 17 have been
   previously associated with this group, but lie outside our survey
   region. (b) The right image shows the velocity distribution for the
   NGC 1566 group. The contour levels for the right image are 600 to
   1615 \kms\ in steps of 35 \kms. The Parkes beam is shown in the
   bottom left corner of each image.}
\label{fig:1566mom}
\end{figure*}

\subsection{Derivation of the HI parameters}

Once the source list was made, the \HI\ parameters for each detection
were derived using {\sc miriad} routines (Sault et al 1995) on the
smoothed cube (velocity resolution 6.6 \kms). A zeroth order moment
map ({\it moment}) was made for the velocity range of emission of each
detection. The central point of emission was then determined by using
{\it imfit} to fit a 2 dimensional Gaussian to this moment map. This
central position was then used in the task {\it mbspect} to make a
spectrum of the source.

A spectrum for each source was produced using a box size of 5 pixels
in width (20 arcmin), assuming a point source. The 20\% and 50\%
velocity widths (taken as the maximum width fit in {\it mbspect}),
integrated flux density (robust moment 0) and peak flux density for each
source were also determined with {\it mbspect}. The heliocentric velocity
for a source was determined to be the centre point of the 20\%
velocity width (in the optical convention).

The integrated flux density was also measured for a box size of 7
pixels to determine if any extended emission surrounded the
detection. Four of the galaxies were found to have extended
emission. These four sources were fitted in a similar way to above,
but in this case, the flux within the box was summed rather than
assuming a point source, and the box size was iteratively increased
until the total flux density was constant. The final parameters for
all galaxies detected in \HI\ are given in Table~\ref{tab:hi_params},
including the fit parameters for the spectra. The \HI\ mass of each
detection was determined using \MHI = $2.356 \times 10^5 D^2 F_{HI} $,
where $F_{HI}$ is the integrated \HI\ flux density in Jy \kms\ (see
Table~\ref{tab:hi_params}), and D is the distance to the group in Mpc
(21 Mpc for the NGC 1566 group).

\HI\ spectra for each of the galaxies are shown in
Figure~\ref{fig:hispectra}. These spectra show the fitted baseline,
the region excluded from the baseline fit with the vertical lines, and
the 20\% and 50\% velocity widths are shown by the crosses (minimum)
and circles (maximum). The peak flux density is shown with a filled
circle.

Uncertainties on the derived parameters were calculated according to
Koribalski et al. (2004). Detections with low \HI\ flux have large
uncertainties in the determined velocity, as do detections with asymmetric
profiles. The uncertainty on the peak flux density is always greater
than the rms noise of the fitted cube. Positional uncertainties are
not quoted. However, the positional uncertainty of the \HI\ detections
can be calculated as the gridded beamsize divided by the
signal-to-noise (Koribalski et al. 2004).

\subsection{Optical Identification of the \HI\ Detections}

Optical identification of the \HI\ detections was made using the NASA
Extragalactic Database (NED). A 6 arcmin region around the central
point of the \HI\ emission was searched for previously catalogued
galaxies of either the same redshift or unknown redshift in
NED. Table~\ref{tab:opt_hi} gives the identifications for the \HI\
detections, with their previously determined velocities. Eleven \HI\
sources in the NGC 1566 group had matching optical counterparts. Of
these, two \HI\ sources (\#5 and \#11) correspond to catalogued galaxy
pairs.  The remaining two \HI\ detections had no previously catalogued
galaxy of known redshift nearby, and further high resolution
observations were necessary to determine the optical counterparts to
these sources (see Section 3 for details). Optical images for each
detection were downloaded from the Digital Sky Survey\footnote{The
Digital Sky Survey, provided by the Space Science Institute, based on
photographic data from the UK Schmidt Telescope.}  for each of the
galaxies. These can be seen in Figure~\ref{fig:dssimages}. The DSS
images are 10 arcminutes square, and are in $R$-band. A variety of
galaxy types can be seen from bright spirals and elliptical galaxies
to small, faint dwarf galaxies. The optical characteristics for all
galaxies are given in Table~\ref{tab:opt_hi}.

\section{Results} 
\subsection{\HI\ Properties of the NGC 1566 Group}

\begin{figure*}
\begin{tabular}{cc}
\mbox{\psfig{file=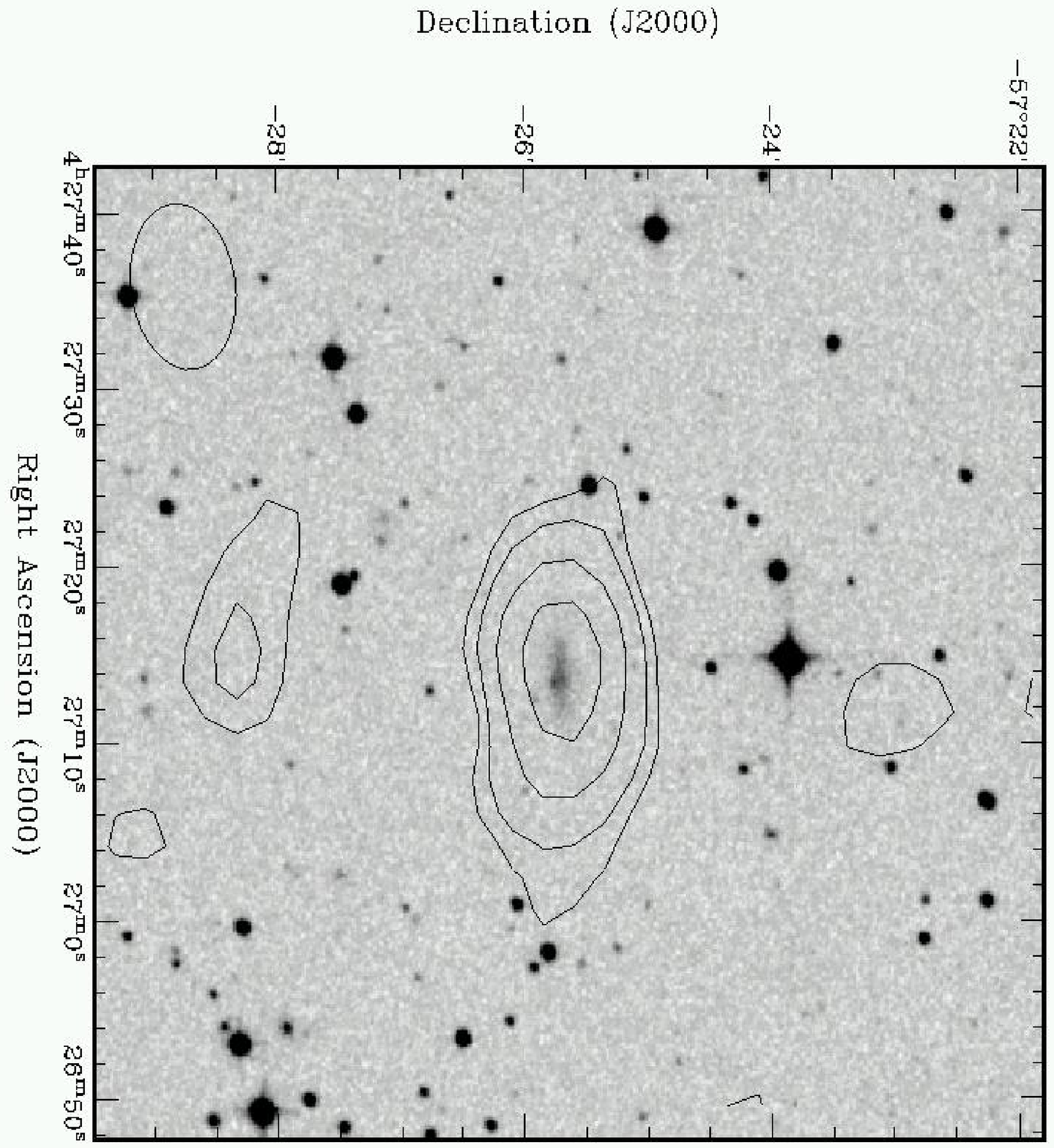,width=8cm}} &
\mbox{\psfig{file=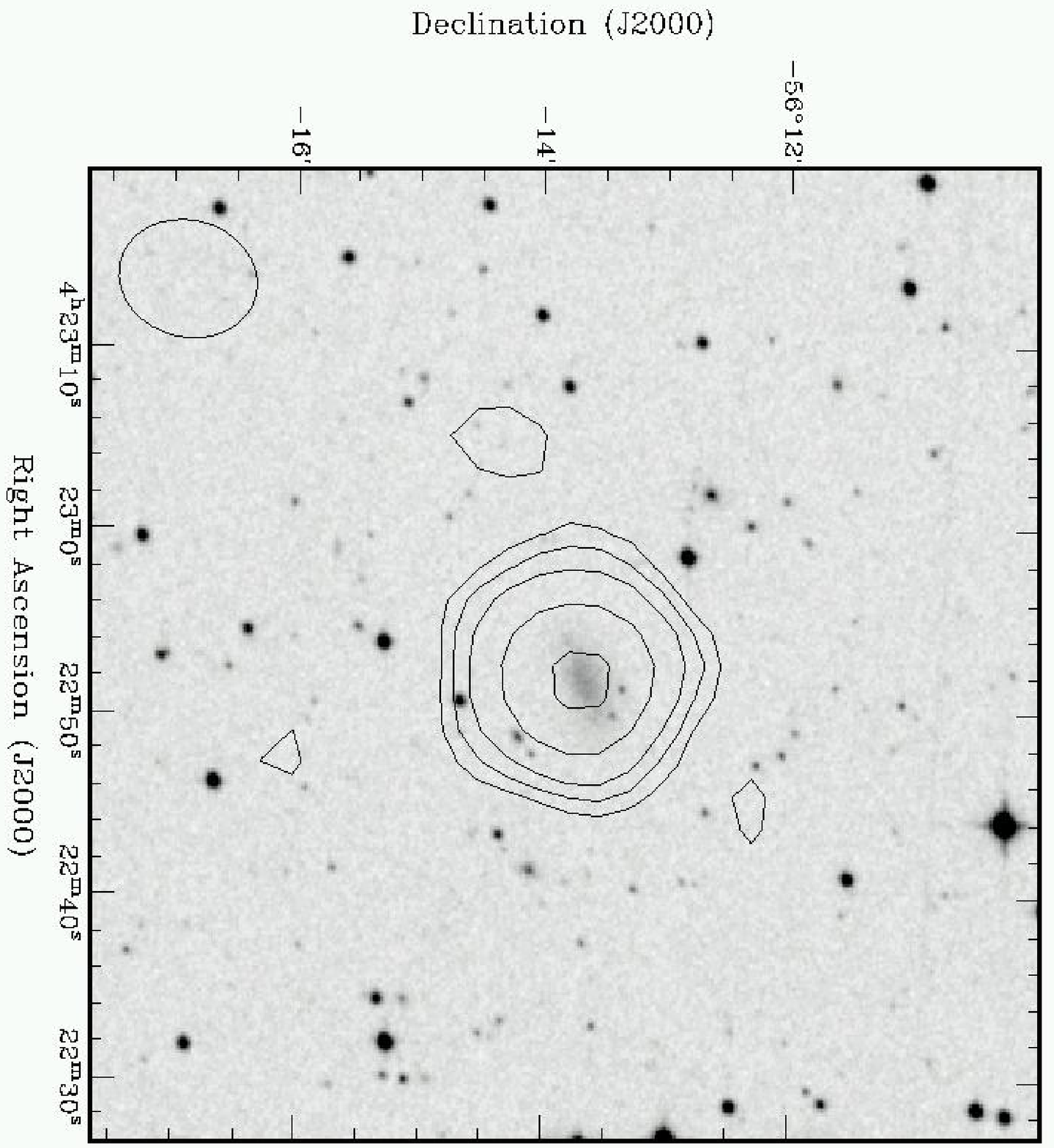,width=8cm}}\\ 
\end{tabular}
\caption{ATCA \HI\ distribution overlaid on DSS II $R$-band images of the two
new group members of the NGC 1566 group - LSBG F157-081 (left) \&
APMBGC 157+016+068 (right). The contour levels are 0.15, 0.25, 0.4,
0.75, 1.25 Jy \kms\ (starting at 0.25 Jy beam$^{-1}$ \kms\ for the left
image). The ATCA beam is shown in the bottom left corner of each image. }
\label{fig:ATCA_maps}
\end{figure*}

The \HI\ distribution for the NGC 1566 group is shown in
Figure~\ref{fig:1566mom}(a), and the mean \HI\ velocity field is
shown in Figure~\ref{fig:1566mom}(b). These were derived from the
narrow-band data cubes. After Hanning smoothing the data to the same
spectral resolution as HIPASS, we used the AIPS task {\sc momnt} (same
smoothing parameters as for
Figure~\ref{fig:hipmom0}) to derive these \HI\
moment maps. Because of varying baseline curvature (see
Fig.~\ref{fig:hispectra}) these moment maps are not as sensitive as
the individual \HI\ spectra. We used a flux density cutoff of 10
mJy\,beam$^{-1}$.


We detected 13 sources in the \HI\ datacube, compared with 24 optical
galaxies with known velocities in the region from LEDA. Two of the
\HI\ sources correspond to interacting galaxy pairs. We also
determined the redshift for two galaxies previously not known to be
group members, thus taking the number of known members for this group
to 26. The spatial distribution of all known galaxies in the NGC 1566
group is over-plotted on the \HI\ distribution in
Figure~\ref{fig:1566mom}a. The crossed symbols are those detected in
\HI\ and open symbols were not detected in our survey. The ovals
represent late-type galaxies, and the circles are E/S0 galaxies. The
\HI\ distribution shows an apparent connected feature encompassing
galaxies \#2, \#5a, \#5b , \#6, \#7, \#8 and \#20. However, except for around the
interacting system of NGC 1533/IC 2038, it is likely that the this
apparent feature is due to smearing of the beam in the \HI\ dataset,
and higher resolution \HI\ observations of the region are needed
determine the nature of the \HI\ distribution.

Nine known members of the NGC 1566 group were not detected in our \HI\
survey (to an \HI\ mass limit of $\sim 3.5 \times 10^8 $ \Msun), and
two further members were outside the survey region. We detected four
E/S0 galaxies in \HI\, but did not detect seven other early-type
galaxies previously determined to be in the group. There was one
irregular and one Sa galaxy not detected in \HI\, but all other known
spiral/late-type galaxies were detected in \HI. NGC 1553 has
previously been reported as detected in \HI\ (de Vaucouleurs et
al. 1991), however the resolution of our Parkes data is too coarse to
determine a separate detection for this galaxy.

The two new group members were observed at the Australia Telescope
Compact Array from 2004 March 3-7. The data were reduced using
standard MIRIAD routines, and the resulting \HI\ distributions are
shown in Figure~\ref{fig:ATCA_maps}. Both of the \HI\ detections
correspond to previously catalogued low surface brightness irregular
galaxies with blue luminosities around $10^8 L_{\odot}$. These two
galaxies are LSBG F157-081 \& APMBGC 157+016+068 (as named in NED),
and both have no previous known redshift. The \HI\ mass (as determined
from the Parkes data) for the two galaxies is $3.3 \times 10^8 $
\Msun\ and $2.5 \times 10^8 $ \Msun\ respectively. Both of these new
group members were unresolved in the ATCA \HI\ maps.

\begin{figure*}
\centering
\begin{tabular}{cc}
\mbox{\psfig{file=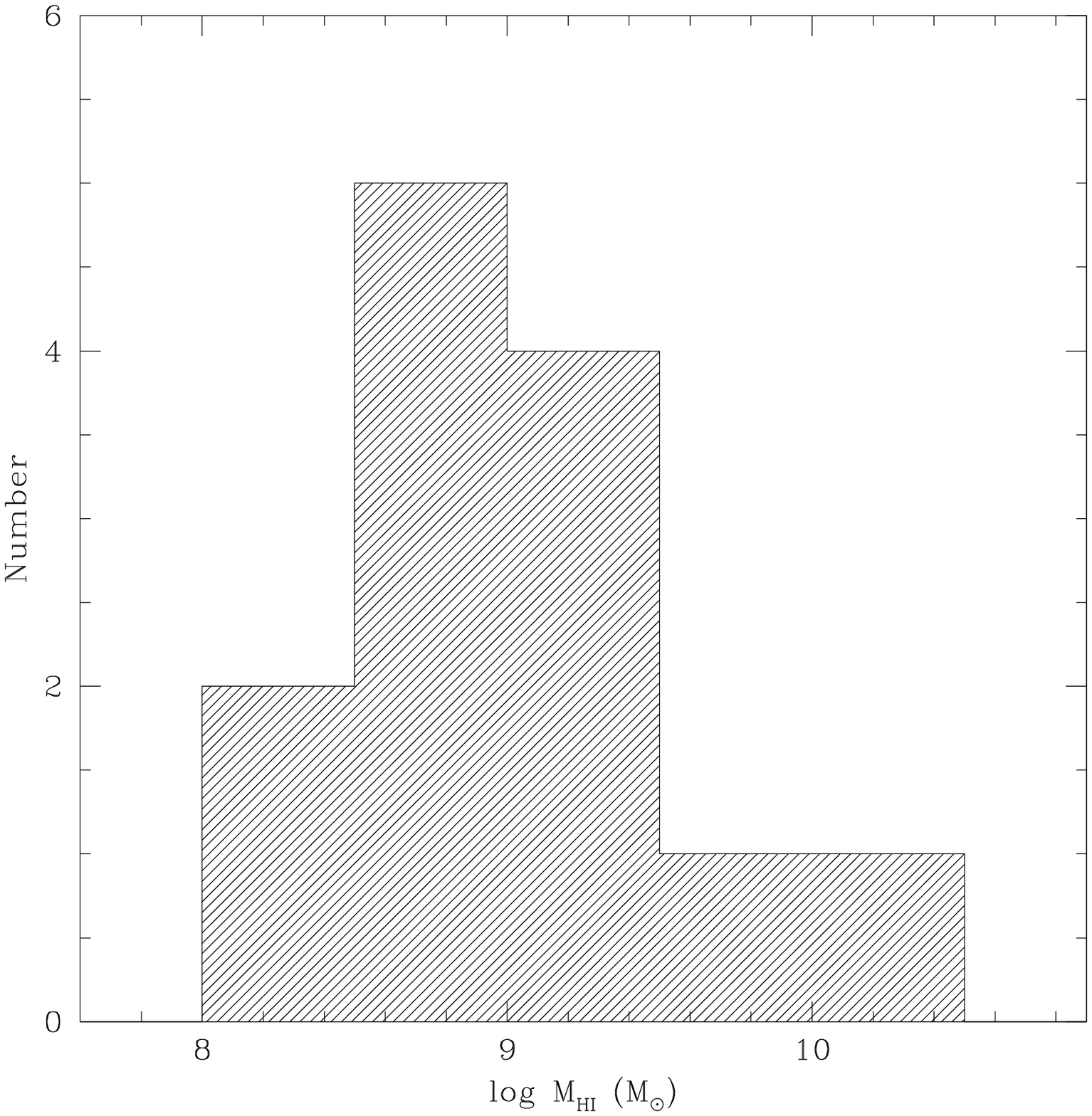,width=7.5cm,angle=0}}&
\mbox{\psfig{file=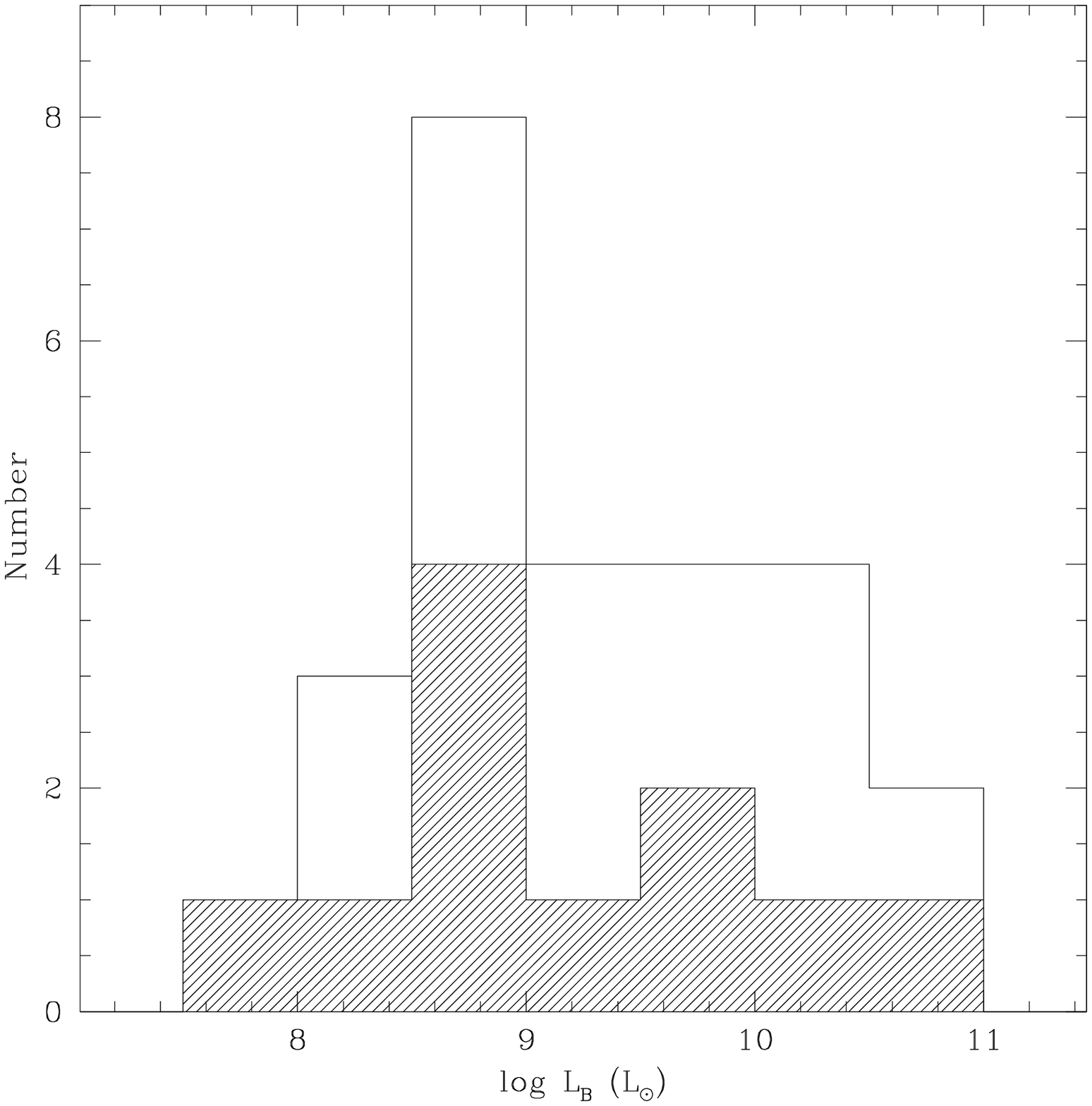,width=7.5cm,angle=0}}\\
\end{tabular}
\caption{ \HI\ mass distribution for NGC 1566 group (left). Luminosity
distribution for known members of the group (right). The open histogram
indicates all members known optically, and the filled histogram
indicates those galaxies detected in \HI. }
\label{fig:hi_hist}

\end{figure*}

The \HI\ mass of the sources we detected in the Parkes data range from
$2.1 \times 10^8 $ \Msun\ to $1.4 \times 10^{10} $ \Msun (see
Figure~\ref{fig:hi_hist}). The highest \HI\ mass galaxy was NGC 1566
itself. This galaxy is extended in the Parkes beam. We detect an
integrated \HI\ flux density of $148.1 \pm 6.4$ Jy \kms, which is
consistent with previously measured values. The HIPASS Bright Galaxy
Catalogue (BGC; Koribalski et al. 2004) measured $140 \pm 10$ Jy \kms,
as do Mathewson \& Ford (1996). Walsh (2004) measured $147 \pm 10$ Jy
\kms\ and $161 \pm 15$ from Parkes and the ATCA respectively for NGC
1566. The total \HI\ mass we detected in the NGC 1566 group is 3.5
$\times 10^{10} $ \Msun, thus the galaxy NGC 1566 itself accounts for
nearly half the \HI\ in the group.  We found that over half the
galaxies in this group have \HI\ masses less than $10^9 $ \Msun.


\subsection{Optical Properties of the NGC 1566 group}

The total $B$-band magnitudes for all known members in the NGC 1566
group are listed in Table~\ref{tab:opt_hi}. Total luminosities were
determined from these magnitudes (after the magnitudes were corrected
for Galactic extinction (Schlegel et al. 1998)), and are also listed in
this table. The highest luminosity galaxy in the NGC 1566 group is NGC
1553, with a luminosity of $4 \times 10^{10}L_{\odot}$. NGC 1553 is an
S0 galaxy that was undetected in our \HI\ survey. The second brightest
galaxy in the group is NGC 1566 with $L_B=3.7 \times
10^{10}L_{\odot}$. The two lowest luminosity galaxies are the two new
members detected in our \HI\ survey, LSBG F157-081 and APMBGC
157+016+068, with luminosities of 9 $\times 10^7 L_{\odot}$ and
1.3 $\times 10^8 L_{\odot}$ respectively.

The \HI\ mass to $B$-band luminosity ratio (\MHI/$L_B$) for the
galaxies in the NGC 1566 group varies from $< 0.01$ for the early type
galaxies NGC 1553 and NGC 1549, to 3.7 for the dwarf irregular galaxy
LSBG F157-081. The total $B$-band luminosity for the group is 1.8
$\times 10^{11}\,L_{\odot}$, compared to the total \HI\ mass of the
group of 3.5 $\times 10^{10}$ \Msun. This gives an overall \MHI/$L_B$
of 0.19 \Msun/$L_{\odot}$. The two new group members have \MHI/$L_B$
of 3.7 and 1.9, which confirms \HI\ surveys are a good way of
detecting gas-rich, low luminosity group members.

Using the $B$-band magnitudes, we have constructed a luminosity
histogram for this group. This is shown in
Figure~\ref{fig:hi_hist}. Shown is the total luminosity distribution
for the group, and that for only those galaxies detected in \HI\ (NB:
only single \HI\ detections are included in the second sample). We
find that the \HI\ detected galaxies span the full range of the
optical luminosity distribution, and have a similar shape to the total
luminosity distribution. Similar to the \HI\ mass distribution for
this group, half of the galaxies in the group have
$L_B<10^9L_{\odot}$. The optical galaxy numbers drop off
towards $L_B$ of $10^8L_{\odot}$, which corresponds to the optical
limit for previous redshift surveys of the region.

\begin{table*} 
\caption{Group members of the NGC 1566 group. Galaxies 1-13 were
detected in the \HI\ survey, and galaxies 14-24 have optical
detections only.}
\label{tab:opt_hi} 
\begin{tabular}{llrlcrl}
\hline
Galaxy No. &  Galaxy Name    &Velocity & Morphology & $B$-band magnitude  & $L_B$ &\MHI/$L_B$\\
(1) &   (2)            & (3)       &(4)      &(5)     &(6)    &(7) \\
\hline
1  & IC~2049$^1$           &1086$\pm46$ & Sd        & 14.56 &   6.6&0.9 \\
2  & NGC~1536          &1565$\pm104$& Sc pec    & 13.29 &  21.3&0.1\\
3  & NGC~1543          &1094$\pm32$ & S0        & 10.67 & 237.9&0.04 \\
4  & LSBG~F157--081    &$\cdots$    & Irr       & 16.72 &   0.9&3.7\\
5a  & NGC~1533         &668$\pm41$  & S0        & 11.74 &  88.8&$\cdots$\\
5b  &IC~2038           &712$\pm52$  & Sd pec    & 14.82 &   5.2&$\cdots$\\
6 & APMBGC~157+016+068 &$\cdots$    & Irr       & 16.32 &   1.3&1.9\\
7 & NGC~1546           &1160$\pm66$ & S0        & 12.31 &  52.5&0.5\\
8 & IC~2058            &1268$\pm88$ & Sc        & 13.90 &  12.1&1.4\\
9  & IC~2032           &1070$\pm34$ & Im        & 14.78 &   5.4&1.1\\
10 & NGC~1566          &1449$\pm19$ & Sbc       & 10.19 & 370.2&0.4\\
11a & NGC~1596         &1465$\pm24$ & S0        & 12.01 &  69.2&$\cdots$\\
11b & NGC~1602         &1731$\pm31$ & Im        & 13.79 &  13.4&$\cdots$\\
12  & NGC~1515         &1216$\pm48$ & Sbc       & 11.96 &  68.5&0.2\\
13  & NGC~1522         &1012$\pm91$ & pec       & 14.08 &  10.3&0.5\\
\hline		         	       	   	    	 
14&ESO~118-019$^2$         &1133$\pm77$ & S0 pec    &  14.90 &   4.8&$<0.7$\\
15&ESO~157-030         &1341$\pm51$ & E4        &  14.30 &   8.4&$<0.5$\\
16&ESO~157-047$^{3,4}$     &1733$\pm39$ & S0/a pec  &  15.52 &   2.7&$<1.29$\\
17&ESO~157-049$^3$     &1729$\pm34$ & Sbc       &  14.37 &   7.9&$<0.4$\\
18&IC~2085             &1010$\pm30$ & S0 pec    &  14.26 &   8.7&$<0.4$\\
19&NGC~1549            &1214$\pm21$ & E0-1      &  10.48 & 283.4&$<0.01$\\
20&NGC~1553$^5$            &1239$\pm18$ & S0        &  10.10 & 402.2&$<0.01$\\
21&NGC~1574            & 925$\pm66$ & S0        &  11.17 & 150.1&$<0.02$\\
22&NGC~1581            &1527$\pm88$ & S0        &  13.36 &  20.0&$<0.18$\\
23&NGC~1617            &1063$\pm21$ & Sa        &  11.35 & 127.2&$<0.03$\\
24&PGC~429411          &1135$\pm40$ & Irr       &  16.19 &   1.5&$<2.3$\\
\hline
\end{tabular}

\flushleft The columns are: (1) GEMS galaxy number, (2) galaxy name,
(3) optical velocity (km s$^{-1}$) from the RC3 catalogue (de
Vaucouleurs et al. 1991), except where listed in the table notes
below, (4) morphological classification, (5) $B$-band magnitude, (6)
$B$-band luminosity in units of $10^8 L_{\odot}$, (7) \MHI/$L_B$ in
units of \Msun/$L_{\odot}$, where no value is given for the
interacting galaxies, and the upper limits are based on the \HI\ mass
limit of the survey of $3.5 \times 10^8$ \Msun. The $B$-band
magnitudes come from the ESO-LV catalogue (Lauberts \& Valentijn
1989), apart from \#4 and \#6, which come from Maddox et al. (1990),
and \#24 which is the extinction corrected total magnitude from the
LEDA database. \\\smallskip Notes: 1. The \HI\ velocity for this
galaxy is $1466\pm3$ \kms, which is inconsistent with the published
optical value, 2. \#14 Velocity from Saunders et al. (2000), 3. These
galaxies have previously been associated with the NGC 1566 group but
lie outside our \HI\ survey region, 4. \#16 Velocity from Loveday et
al. (1996), 5. \#24 Velocity from Katgert et al. (1998).
\end{table*}

\section{Discussion}
\subsection{\HI\ Content and Deficiency}

\HI\ observations of galaxies in groups and clusters have shown some
galaxies to be \HI\ deficient near the centre of the group or cluster
(Solanes et al. 2001; Verdes-Montenegro et al. 2001; Giovanelli \&
Haynes 1985).  In order to determine whether the \HI\ content of a
galaxy in a group or cluster is deficient, normal or over-abundant, a
good estimate of the expected \HI\ is needed. The expected \HI\ mass
of isolated galaxies has been studied in the past using either the
optical morphology and optical diameter of the galaxies (Solanes et
al. 1996; Chamaraux et al. 1986; Haynes \& Giovanelli 1984 [hereafter
HG84]) or using blue magnitude and optical morphology (HG84). HG84
show that the \HI\ content is more closely tied to the galaxy's
optical extent and morphological type, thus we use this method in the
following analysis.

Optical properties for most previously catalogued galaxies in the
HIPASS BGC (Koribalski et al. 2004) were obtained from LEDA, including
morphological type, optical diameters and magnitudes (Koribalski et
al. 2004). Following Solanes et al. (1996), we made a linear
regression fit for optical diameter and \HI\ mass to nearly 800 BGC
galaxies.  The galaxies were divided into morphological type from Sb
to Irr, and the resulting linear regression coefficients, $\alpha$ and
$\beta$ (such that $M_{HI,ex} = \alpha + \beta\, {\rm log }d$, where
$d$ is the optical diameter of the galaxy in kpc) for each
morphological type are listed in Table~\ref{tab:lin_reg}. The scatter
in the \HI\ masses, $\sigma$, and the number galaxies at each
morphological type, N, are also given in Table~\ref{tab:lin_reg}. We
use the morphology and optical diameter of the NGC 1566 group
galaxies, and the coefficients given in Table~\ref{tab:lin_reg}, to
determine their expected \HI\ mass, $M_{HI,ex}$. There are very small
numbers of early-type galaxies in the BGC, thus there is a large error
on calculating the expected \HI\ mass for galaxies of optical
morphology earlier than Sb. In particular, it is known that elliptical
galaxies are rarely detected in \HI\ (Sadler et al. 2001), and we do
not discuss these galaxies further in this section.

\begin{table} 
\centering
\caption{Linear regression coefficients as derived from the HIPASS
BGC (Koribalski et al. 2004).}
\label{tab:lin_reg} 
\begin{tabular}{lcccc}
\hline
Type   & $\alpha$ & $\beta$ & $\sigma$& N \\
\hline
Sb               &  8.39 & 1.02 & 0.24 &     85        \\
Sbc              &  8.39 & 1.05 & 0.29 &     130       \\ 
Sc               &  8.30 & 0.99 & 0.27 &     134       \\
Scd              &  7.80 & 1.38 & 0.24 &     88        \\
Sd               &  7.12 & 1.97 & 0.26 &     75        \\
Sm               &  7.70 & 1.45 & 0.28 &     84        \\
Irr              &  7.29 & 1.9  & 0.33 &     141       \\
\hline
\end{tabular}
\end{table}

We believe that the \HI\ estimates from the BGC for late-type
galaxies, especially Sd-Irr are more accurate than previous
estimates. Low surface brightness late-type galaxies can be easily
missed in optical surveys, and thus these potentially \HI-rich
galaxies will not have been included in a targeted \HI\ survey. As
late-type galaxies have a high detection rate in \HI, we expect to
have a more complete sample than optically based pointed \HI\
surveys. Indeed, the latest \HI\ estimate for late-type galaxies in
the literature was HG84 who have a sample size of just 7 for irregular
galaxies, whereas there are 141 in the BGC. We must bear in mind that
we have no indication of how many late-type galaxies were not detected
in the survey, thus the upper limit to the \HI\ mass of undetected
galaxies is not included in the linear regression fit.

\begin{figure}
\mbox{\psfig{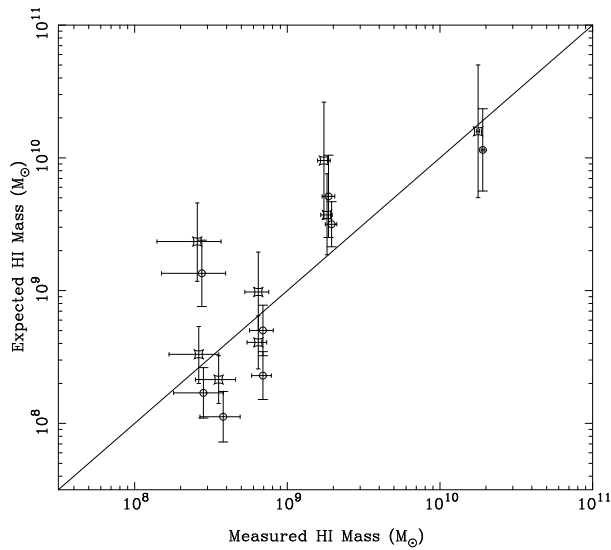}} 
\caption{Expected \HI\ mass of non-interacting late-type galaxies
compared to the detected \HI\ mass. The squares show the
expected \HI\ mass using the coefficients in Table~\ref{tab:lin_reg},
and the circles are the expected \HI\ mass from the coefficients
in HG84. The errors show the scatter in \HI\ mass for a galaxy of
given optical diameter and morphology. }
\label{fig:hi_def}
\end{figure}

We have calculated the expected \HI\ mass for each galaxy in the NGC
1566 group, including previously catalogued members of known redshift
that were not detected in our \HI\ survey. To reduce the uncertainty
in the calculation of expected \HI\ mass, we limit our comparison to
non-interacting galaxies of type Sb or later. Nine galaxies meet this
criteria, including eight galaxies detected in our \HI\ survey, and
the undetected irregular galaxy PGC 429411. For comparison we
calculate the expected \HI\ mass using both the coefficients in
Table~\ref{tab:lin_reg}, and those given in
HG84. Figure~\ref{fig:hi_def} shows the expected \HI\ mass versus the
detected \HI\ mass for the eight galaxies detected in our survey. The
vertical error bars represent the scatter in the expected \HI\ mass of
a given \HI\ mass and optical diameter for the BGC and HG84
samples. This figure shows that the expected \HI\ mass from HG84 is
always lower than that from the BGC.

The expected \HI\ mass for most of the galaxies is consistent with
what was observed, and considering the group as a whole, the NGC 1566
group does not appear to be \HI\ deficient.  Two galaxies have a
slightly higher detected \HI\ mass than was expected, and a further
two galaxies display a marked \HI\ deficiency.  These latter galaxies
are the Sbc spiral NGC 1515 and the peculiar Sc spiral NGC 1536, with
$5-10$ times less \HI\ detected than expected. While NGC 1515 is
relatively isolated, NGC 1536 is situated spatially near the
interacting galaxy pair of NGC~1533/IC~2038. However there is a large
velocity difference between the pair and NGC 1536 of nearly
500\kms. The expected \HI\ mass for the undetected irregular galaxy
PGC 429411 is $3.1 \pm 1 \times 10^8 $ \Msun. As this \HI\ mass lies
near our detection limit for the \HI\ survey, we are unable to tell
whether this galaxy is \HI\ deficient, or contains \HI\ within the
normal range.

\subsection{Group Distribution and Dynamics} 

\begin{figure} 
\mbox{\psfig{file=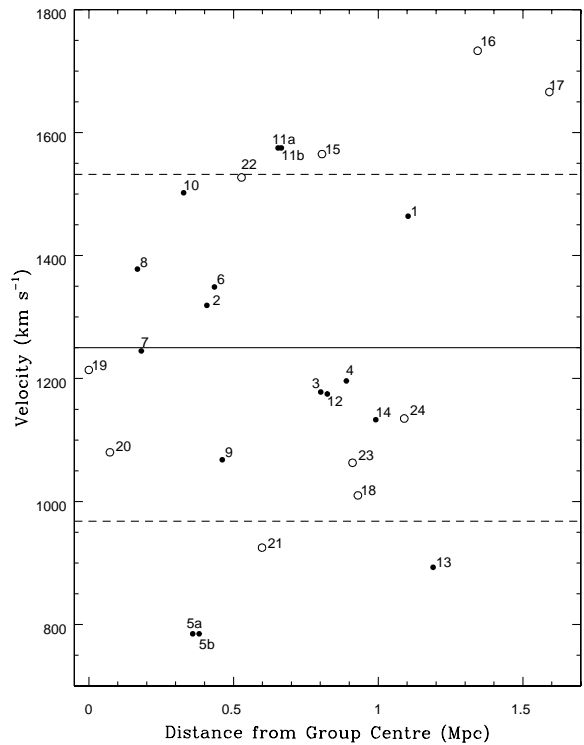,width=8cm}}\\ 
\caption{Velocity of the NGC 1566 group members versus their distance
from group center. The filled circles show galaxies detected in \HI\,
and the open symbols are known galaxies in the group that we did not
detect. The solid line shows the mean velocity for the group, and the
dashed lines show the velocity dispersion. The virial radius for this
group is 580 kpc. Galaxy \#19 (NGC 1549) is taken as the centre of the
group. }
\label{fig:vel_dist}
\end{figure}

To study the dynamical state of the NGC 1566 group, we now derive the
virial radius and velocity dispersion of the group members. The virial
radius can be calculated using X-ray emission where the latter exists
in the intra-group medium.  The ROSAT PSPC data for the NGC~1566
group, along with other GEMS groups, are presented in OP04. The X-ray
data cover a circular region of $\sim$ 2 degrees in diameter. OP04 find
X-ray emission that is consistent with coming from NGC~1566 itself,
i.e. no evidence for extended emission from the intra-group
environment.

OP04 derived an $r_{500}$ radius of 470 kpc based
on the X-ray temperature of 0.70 $\pm$ 0.11 keV. However, the X-ray
emission in the NGC~1566 group is dominated by hot gas in the halo of
NGC~1566 itself, which reaches the background level at a radius of 29
kpc. So the temperature-based $r_{500}$ should be treated with some
caution.

An alternative method for calculating $r_{500}$ is to use the group's
velocity dispersion and the virial theorem (see OP04). The accuracy of
this method improves with the number of reliable galaxy velocities.
Based on only 9 galaxies, OP04 derive a velocity dispersion for the
group of 184 $\pm$ 47 km/s. This would translate into an $r_{500}$
radius of 250 kpc.

A better estimate of the dispersion for the group uses all known group
members. Velocities were taken from the literature for the optically
known galaxies not detected in our \HI\ survey. Without removing any
spatial or velocity outliers, we use a biweight estimator with
bootstrap errors as per Beers et al. (1990). This gives the average
velocity of the group as $v=1250 \pm 57$ \kms, and velocity dispersion
of $\sigma_v=282\pm30$ \kms. This velocity dispersion gives a
$r_{500}$ radius of 386 kpc, and thus a virial radius, $r_{\rm v}$, of
$\sim$580 kpc for the NGC 1566 group ($r_{\rm v}$ $\sim$ 3/2 $\times$
$r_{500}$). 

Figure~\ref{fig:vel_dist} shows the velocity distribution versus the
distance from the centre of the group. The solid line and dashed lines
show $v$ and $\sigma_v$ respectively.  For this plot, the galaxy \#19,
NGC 1549, was chosen to be the center position as it is the brightest
elliptical in the NGC 1566 group, and has a velocity very close to the
mean velocity of the group.

Eleven galaxies associated with the NGC 1566 group lie within the
virial radius of 580 kpc. A further 12 galaxies are located between
$1-2$ $r_{\rm v}$. As such a large number of galaxies lie beyond the
virial radius, it is possible this group is young and not yet
virialised. There has been recent work on the evolution of galaxies
lying outside the virial radius of groups and clusters. Simulations
have shown that up to half of the galaxies that lie beyond the virial
radius of a cluster may have traveled through the cluster center
(Gill, Knebe \& Gibson 2004; Ghigna et al. 1998). Thus the
evolutionary history of these 'backsplash' galaxies may have been
influenced by this encounter, and mass loss of up to 40\% or greater
might be expected. Observationally, there is some evidence of \HI\
stripped galaxies not only in the centres of clusters (e.g. Solanes et
al. 2001), but also nearer the edges of clusters (Kenney, van Gorkom
\& Vollmer 2004; Vogt et al. 2004; Vollmer 2003). By determining the
\HI\ content of the galaxies in the GEMS groups we will be able to see
if this trend continues in the less dense group environment. We have
found two \HI\ deficient galaxies in the NGC 1566 group, NGC 1536 and
NGC 1515, which lie near or beyond the virial radius of the
group. High resolution \HI\ observations of these two \HI\ deficient
galaxies combined with H$\alpha$ measurements (e.g.  Vogt et al. 2004)
will provide further information to the past history of these
galaxies, particularly whether their gas removal mechanism is due to
tidal interactions, ram pressure stripping, or a combination of the
two.

\section{Conclusions} 

We have conducted a blind \HI\ survey of a region of 5.5\degr\
$\times$ 5.5\degr\ in the NGC 1566 group. Thirteen galaxies were
detected in the \HI\ datacube, including two LSB dwarf galaxies that
had no previously known redshifts (LSBG F157-081 \& APMBGC
157+016+068), which are new group members. There are now 26 known
members in the NGC 1566 group. Two \HI\ detections have two or more
optical counterparts, and are previously known interacting systems
that are confused within our beamsize of 15.5 arcmin (NGC 1533/IC
2038, and NGC 1602/NGC 1596). No isolated \HI\ clouds were detected in
our survey to a limit of $\sim 3.5\times 10^8 $ \Msun.

The total \HI\ mass detected in the group was $3.5 \times 10^{10}$
\Msun, and the galaxy NGC 1566 itself contains nearly half of the \HI\
mass of this group, with \MHI = $1.4 \times 10^{10}$ \Msun.

 We calculate the virial radius of this group to be 580 kpc, and find
over half the galaxies we associate with the group to lie beyond this
radius. Thus it is possible that this is a young, non-virialised
group. The total \HI\ contained in the late-type galaxies of this
group appears to be consistent with the expected \HI\ content based on
their optical diameter and morphology. However there are two cases in
which a spiral galaxy appears to be 5-10 times more deficient than
expected - these are NGC 1536 and NGC 1515. Further observations are
needed of these two galaxies to determine their gas removal mechanisms.

\section*{Acknowledgments}

Thanks to C. Mundell, N. McKay and S. Brough for help with the Parkes
observations, and to R. Allen for assistance with the ATCA
observations.  We are grateful to D. J. Pisano and R. Braun for useful
discussions relating to the data reduction, in particular ideas for
source masking during bandpass estimation to improve dynamic range
near bright sources. J. Osmond and T. Ponman are acknowledged for
helpful discussions.  We acknowledge the Parkes telescope staff and
thank them for their assistance in the observations. Many thanks to
M. Calabretta for continuing support of livedata. Thankyou to the
anonymous referee for helpful comments.

This research has made extensive use of the NASA/IPAC Extragalactic
Database (NED) which is operated by the Jet Propulsion Laboratory,
Caltech, under contract with the National Aeronautics and Space
Administration. The Digitized Sky Survey was produced by the Space
Telescope Science Institute (STScI) and is based on photographic data
from the UK Schmidt Telescope, the Royal Observatory Edinburgh, the UK
Science and Engineering Research Council, and the Anglo-Australian
Observatory. VAK acknowledges the support of an ARC/CSIRO Linkage
Postdoctoral Fellowship.


\begin{thebibliography}{}

\bibitem[1995]{bajaja:etal} Bajaja, E., Wielebinski, R., Reuter,
H.-P., Harnett, J. I., Hummel, E. 1995, A\&AS, 114, 147
\bibitem[Banks et al. 1999]{banks99}
Banks, G. D., Disney, M. J., Knezek, P. M. et al. 1999, ApJ, 524, 612
\bibitem[Barnes et al. 2001]{barnes01}
Barnes, D. G., Staveley-Smith, L., de Blok, W. J. G. et al. 2001, MNRAS, 322, 486
\bibitem[Barnes \& Webster 2001]{barneswebster99}
Barnes, D. G., Webster, R. L. 2001, MNRAS, 324, 859
\bibitem[1990]{beers:etal}
Beers, T. C., Flynn, K., Gebhardt, K., 1990, AJ, 100, 32 
\bibitem[2002]{deblok:etal}
 de Blok, W. J. G., Zwaan, M. A., Dijkstra,
M., Briggs, F. H., Freeman, K. C. 2002, A\&A 382, 43
\bibitem[Boyce et al. 2001]{boyce01}
Boyce, P.J., Minchin, R. F., Kilborn, V. A. et al. 2001, ApJL, 560, 127
\bibitem[1993]{carrasco:etal}
Carrasco E.R., Mendes de Oliveira C., Infante L., Bolte M. 2001, AJ 121, 148
\bibitem[1986]{chamaraux:etal}
Chamaraux, P., Balkowski, C., Fontanelli, P. 1986, A\&A, 1986, 165
\bibitem[2004]{chung:etal}
Chung A. et al. 2004, in preparation
\bibitem[2001]{dahlem:etal}
Dahlem, M., Ehle, M., Ryder, S. D. 2001, A\&A 373, 485
\bibitem[1996]{ehle:etal}
Ehle, M., Beck, R., Haynes, R. F. et al. 1996, A\&A, 306, 73
\bibitem[1990]{ferguson:sandage}
Ferguson H.C., Sandage A. 1989, AJ 100, 1
\bibitem[1993]{garcia}
Garcia A.M. 1993, A\&AS 100, 47
\bibitem[1995]{gooch1995} Gooch, R. 1995, in ASP Conf. Se. 77,
Astronomical Data Analysis Software and Systems IV, ed R. A. Shaw,
H. E. Payne, J. E. Haynes (San Fransisco: ASP), 144
\bibitem[1998]{ghigna:etal}
Ghigna, S., Moore, B., Governato, F. et al. 1998, MNRAS, 300, 146
\bibitem[2004]{gill:etal}
Gill, S. P. D., Knebe, A., Gibson, B. K. 2004, MNRAS, submitted
\bibitem[1985]{giovanelli:haynes}
Giovanelli, R. \& Haynes, M. P. 1985, ApJ, 292, 404
\bibitem[1984]{haynes:giovanelli}
Haynes M. P. \& Giovanelli, R. 1984, AJ, 89, 758
\bibitem[1984]{haynes:etal}
Haynes, M. P., Giovanelli, R., Chincarini, G. L. 1984, ARAA, 22, 445
\bibitem[1982]{huchra:geller}
Huchra, J. P., Geller, M. J. 1982, ApJ, 257, 423
\bibitem[1998]{katgert:etal}
Katgert, P., Mazure, A., den Hartog, R., et al. 1998, A\&A Suppl. Ser. 129, 399
\bibitem[2004]{kenney}
Kenney, J. D. P., van Gorkom, J. H., Vollmer, B. 2004, AJ, 127, 3375
\bibitem[1999]{klypin}
Klypin, A., Kravstov, A., Valenzuela, O., Prada, F. 1999, ApJ, 522, 82
\bibitem[2004]{koribalski}
Koribalski, B. et al. in preparation
\bibitem[2004]{koribalski:dickey}
Koribalski, B. \& Dickey, J. M. 2004, MNRAS, 348, 1255
\bibitem[2004]{koribalski:etal}
Koribalski B. {\em et al.} 2004, AJ, 128, 16
\bibitem[1989]{lauberts:valentijn} 
Lauberts, A. \& Valentijn, E. A. 1989, in ``The Surface Photometry
Catalogue of the ESO-Uppsala Galaxies'', Garching Bei Munchen: European
Southern Observatory
\bibitem[1996]{loveday:etal}
Loveday, J., Peterson, B. A., Maddox, S. J., Efstathiou, G. 1996, ApJS, 107, 201
\bibitem[1990]{maddox:etal} 
Maddox, S. J., Sutherland, W. J., Efstathiou, G., Loveday, J. 1990, MNRAS, 243, 692
\bibitem[1989]{maia:etal}
Maia M.A.G., Da Costa L.N., Latham D.W 1989, ApJS 69, 809
\bibitem[1996]{mathewson:etal}
Mathewson D.S., Ford V.L. 1996, ApJS 107, 97
\bibitem[McKay et al. 2004]{mckay2004}
McKay, N. F., Mundell, C. G., Brough, S. et al. 2004, MNRAS, in press 
\bibitem[1999]{moore:etal}
Moore, B., Ghigna, S., Governato, F. et al. 1999, ApJ, 524, 19
\bibitem[1999]{morshidi:etal}
Morshidi-Esslinger Z., Davies J.I., Smith R.M. 1999, MNRAS 304, 297
\bibitem[2004]{osmond:ponman}
Osmond, J.P.F., Ponman T.J. 2004, MNRAS, 350, 1511
\bibitem[1990]{pence:etal}
Pence, W. D., Taylor, K., Atherton, P. 1990, ApJ, 357, 415
\bibitem[2004]{pisano:etal}
Pisano, D. J., Barnes, D. G., Gibson, B. K. et al. 2004, ApJL, 610, 17 
\bibitem[1982]{reif:etal}
Reif K., Mebold, U., Goss, W.M., van Woerden, H., Siegman, B. 1982, 
  A\&A 50, 451
\bibitem[2003]{ryan-weber:etal}
Ryan-Weber, E., Webster, R., Bekki, K. 2003, in ``The IGM/Galaxy Connection:The Distribution of Baryons at Z=0'', ASSL Conference Proceedings, Vol. 281, Edited by J. L. Rosenberg and M. E. Putman
\bibitem[2001]{sadler} Sadler, E. M., 2001, in ``Gas and Galaxy
evolution'', ASP Conf. Se., 240, eds J. E. Hibbard,
M. P. Rupen \& J. H. van Gorkom, 445
\bibitem[Sault et al. 1995]{sault1995} Sault, R. J., Teuben, P. J.,
Wright, M. C. H. 1995, in ASP Conf. Se. 77, Astronomical Data Analysis
Software and Systems IV, ed R. A. Shaw, H. E. Payne, \& J. E. Haynes
(San Francisco: ASP), 433
\bibitem[2000]{saunders:etal}
Saunders, W., Sutherland, W. J., Maddox, S. J. 2000, MNRAS, 317, 55
\bibitem[1998]{schlegel:etal}
Schlegel D.J., Finkbeiner D.P., Davis M. 1998, ApJ 500, 525
\bibitem[1984]{shostak:etal}
Shostak, G. S., Allen, R. J., Sullivan, W. T {\sc iii} 1984, A\&A, 139, 15
\bibitem[2001]{solanes2001:etal}
Solanes, J. M, Manrique, A., Garcia-Gomez, C., et al. 2001, ApJ, 548, 97
\bibitem[1996]{solanes:etal}
Solanes, J. M, Giovanelli, R., Haynes, M. P. 1996, ApJ 461, 609
\bibitem[Staveley-Smith et al. 1996]{staveley1996}
Staveley-Smith, L. et al. 1996, PASA, 13, 243
\bibitem[2004]{stevens:etal}
Stevens, J. B., Webster, R. L, Barnes, D. G., Pisano, D. J., Drinkwater, M. J. 2004, PASA, in press
\bibitem[1987]{tully}
Tully, R. B. 1987, ApJ, 321, 280
\bibitem[1991]{devaucouleurs:etal}
de Vaucouleurs G., de Vaucouleurs A., Corwin Jr. H.G., Buta R.J.,
  Paturel G., Fouqu\'e P. 1991, ``Third Reference Catalogue of Bright 
  Galaxies'' (New York: Springer Verlag), [RC3]
\bibitem[2001]{verdes-montenegro:etal}
Verdes-Montenegro, L., Yun, M. S., Williams, B. A., Huchtmeier, W. K., Del Olmo, A., Perea, J. 2001, A\&A, 377, 812
\bibitem[2004]{vogt04}
Vogt, N. P., Haynes, M. P., Giovanelli, R., Herter, T. 2004, AJ, 127, 3300
\bibitem[2003]{vollmer03}
Vollmer, B. 2003, A\&A, 398, 525
\bibitem[2004]{walsh04}
Walsh, W. 2004, New Astronomy, submitted
\bibitem[1994]{yun}
Yun, M. S., Ho, P. T. P., Lo, K. Y. 1994, Nature, 372, 530
\bibitem[2001]{zwaan}
Zwaan, M. A. 2001, MNRAS 325, 1142
\end{thebibliography}
\end{document}